\documentclass[12pt]{iopart}
\usepackage{graphicx}

\newcommand{\newc}{\newcommand}
\newc{\beq}{\begin{equation}}
\newc{\enq}{\end{equation}}
\newc{\bea}{\begin{eqnarray}}
\newc{\ena}{\end{eqnarray}}
\newc{\D}{\displaystyle}
\newc{\noi}{\noindent}
\newc{\cA}{{\cal A}}
\newc{\cB}{{\cal B}}
\newc{\cC}{{\cal C}}
\newc{\cD}{{\cal D}}
\newc{\cE}{{\cal E}}
\newc{\cN}{{\cal N}}
\newc{\cF}{{\cal F}}
\newc{\cG}{{\cal G}}
\newc{\cH}{{\cal H}}
\newc{\cI}{{\cal I}}
\newc{\cJ}{{\cal J}}
\newc{\cK}{{\cal K}}
\newc{\cL}{{\cal L}}
\newc{\cM}{{\mathbf {M}}}
\newc{\cO}{{\cal O}}
\newc{\cP}{{\cal P}}
\newc{\cPr}{{\cal P}}
\newc{\cQ}{{\cal Q}}
\newc{\cR}{{\cal R}}
\newc{\cS}{{\cal S}}
\newc{\cT}{{\cal T}}
\newc{\mG}{{\mathbf G}}
\newc{\Gg}{{\{\mG\cG\}}}
\newc{\GG}{{\{\mG\mG\}}}
\newc{\Gbarra}{{\{\mG\!\!\!\!\slash\,\}}}
\newc{\gbarra}{{\{\cG\!\!\!\!\slash\,\}}}
\newc{\fbarra}{{f_k}}
\newc{\mF}{{\mathbf F}}
\newc{\mi}{i}
\newc{\sen}{{\rm sen\;}}
\newc{\ra}{\rightarrow}
\newc{\rt}{\right}
\newc{\lt}{\left}
\newc{\sqp}{\phi}
\newc{\cp}{\varphi}
\newc{\csp}{\phi_{cl}}
\newc{\ccp}{\varphi_{cl}}
\newc{\osp}{\overline {\phi}}
\newc{\ocp}{\overline {\varphi}}
\newc{\cj}{{\it j}}
\newc{\sj}{{\mathbf J}}
\newc{\cpmm}{{\varphi_>}}
\newc{\cpm}{{\varphi_<}}
\newc{\spmm}{{\phi_>}}
\newc{\spm}{{\phi_<}}
\newc{\cjmm}{{\cj_>}}
\newc{\cjm}{{\cj_<}}
\newc{\sjmm}{{\sj_>}}
\newc{\sjm}{{\sj_<}}
\newc{\etamm}{{\eta_>}}
\newc{\etam}{{\eta_<}}
\newc{\tcp}{{\tilde {\cp}}}
\newc{\tsp}{{\tilde {\sqp}}}
\newc{\nn}{\nonumber}
\newc{\ob}{\overbrace}
\newc{\ub}{\underbrace}
\newc{\veco}{\Lambda \hat \Omega}
\newc{\bfK}{{\mathbf K}}
\newc{\bfk}{{\mathbf k}}
\newc{\bfq}{{\mathbf q}}
\newc{\bfQ}{{\mathbf Q}}
\newc{\bfP}{{\mathbf P}}
\newc{\bfD}{{\mathbf D}}
\newc{\bfp}{{\mathbf p}}
\newc{\la}{\Lambda}
\newc{\bla}{\overline\Lambda}
\newc{\Ci}{{\rm Ci}}
\newc{\Si}{{\rm Si}}
\newc{\Ei}{{\rm Ei}}
\newc{\poi}{\phantom . \noi}

\begin{document}

\title[Nonequilibrium renormalization group]{A nonequilibrium renormalization group approach to turbulent reheating}

\author{Juan Zanella and Esteban Calzetta}

\address{CONICET and Departamento de F\'{\i}sica, Universidad de Buenos Aires, Ciudad Universitaria, 1428
Buenos Aires, Argentina} \ead{zanellaj@df.uba.ar,
calzetta@df.uba.ar}
\begin{abstract}
We use nonequilibrium renormalization group (RG) techniques to
analyze the thermalization process in quantum field theory, and by
extension reheating after inflation. Even if at a high scale
$\Lambda$ the theory is described by a non-dissipative
$\lambda\varphi^{4}$ theory, the RG running induces nontrivial
noise and dissipation. For long wavelength, slowly varying field
configurations, the noise and dissipation are white and ohmic,
respectively. The theory will then tend to thermalize to an
effective temperature given by the fluctuation-dissipation
theorem.
\end{abstract}

\submitto{\JPA}
The goal of this paper is to show how nonequilibrium
renormalization group (RG) techniques may be applied to study the
thermalization process in quantum field theory and, by extension,
the turbulent reheating period in inflation.

The issue of thermalization  in quantum field theory has received
renewed attention in later years, motivated by applications to
cosmology and to relativistic heavy ion collisions, as well as by
theoretical progress through large scale numerical experiments
\cite{Ber04}. However, there is a lack for analytical, model
robust methods capable of yielding predictions for such
observables as the final temperature and thermalization time
scales. This need is particularly felt in applications to
cosmology, since reheating is more likely than not a complex
phenomena involving several nonlinear fields and the evolving
background geometry \cite{BTW06}.

During inflation, the dominant form of matter in the Universe is a
condensate (the inflaton) which evolves "rolling down the slope"
of its effective potential. When the inflaton nears the bottom of
the potential well, it begins to oscillate and transfers its
energy to ordinary matter (then in its vacuum state). We call this
process reheating. Reheating proceeds through several stages
\cite{FeKo06} correlated with the different stages of the
thermalization process, namely preheating, inflaton fragmentation
and turbulent thermalization. Generally speaking,  the early
phases produce an spectrum with high occupation numbers in a
narrow set of modes. Turbulent thermalization concerns the spread
of the spectrum over the full momentum space and the final
achievement of a thermal shape.

At early times  occupation numbers are high and the process may be
described in terms of classical wave turbulence \cite{ZLF92}. As the spectrum
spreads occupation numbers fall and the classical approximation
breaks down. The challenge for us is to provide a quantum
description of turbulent reheating.

For demonstration  processes, we shall only  discuss quantum
turbulent thermalization in a nonlinear scalar field theory in
$3+1$ flat space-time.

The basic idea is  the same as in Kolmogorov - Heisenberg
turbulence theory: a mode of the field with wave number $k$ lives
in the environment provided by all modes with wave number $k' >
k$. Since the physical mechanism for damping in the long
wavelength sector is the interaction with shorter wavelength
modes, it is natural to understand damping as a feature of the
effective theory where the shorter modes have been coarse-grained
away \cite{LOMA96,CHM}. Since this operation will leave the long
wavelength modes in a mixed state, the natural description of the
relevant sector is in terms of a density matrix \cite{Pol06}, and
the natural action functional encoding the effective dynamics is
the Feynman-Vernon Influence Functional (IF) \cite{if1, if2}. Now
suppose we are given the IF when all modes $k>\Lambda$ have been
coarse-grained away, and we wish to further coarse grain the modes
in the range $\Lambda\ge k > k_{0}$. We split the desired range
into shells of infinitesimal thickness $\delta s$, and integrate
out each shell retaining only terms of order $\delta s$. Adding a
change of units and a rescaling of the fields after each
integration, we transform the shell coarse-graining into a RG flow
in the space of influence functionals \cite{WiKo74}. Because we
shall not assume equilibrium conditions, this may be called the
nonequilibrium RG.

Here $\Lambda$ is not meant as an ultraviolet cutoff to be removed
eventually, but rather as the ``hard'' scale at which the
microscopic theory is well understood and radiative corrections
are perturbative. Our goal is to investigate physics at ``soft''
scales $k_{0}\ll\Lambda$. This issue has been studied in the
context of hot nonabelian plasmas, where the emergence of
dissipation and noise has been demonstrated in different ways
\cite{BOD98,BOD99,LIMA02}. We wish to emphasize the nonequilibrium
aspects of the problem, as well as to put those findings on a more
general base by adopting the RG approach.

It is important to stress  two basic differences between the
nonequilibrium and equilibrium RG \cite{LIT98}. The IF may be regarded as an
action for a theory defined on a ``closed time path'' (CTP)
composed of a first branch (going from the initial time $t=0$ to a
later time $t=T$ when the relevant observations will be performed
-that is why we need the density matrix at $T$) and a second
branch returning from $T$ to $0$ \cite{ctp1, ctp2, ctp3, ctp4}.
Thus each physical degree of freedom on the first branch acquires
a twin on the second branch -we say the number of degrees of
freedom is doubled. The  IF is not just a combination of the usual
actions for each branch, but also admits direct couplings across
the branches. The damping constant $\kappa$ and the noise constant
$\nu$ are associated to two of those ``mixed'' terms. Therefore,
the structure of the IF (from now on, CTP action, to emphasize
this feature) is much more complex than the usual Euclidean or
``IN-OUT'' action.

The second fundamental  difference is the presence of the
parameter $T$ itself. In nonequilibrium evolution, it is important
to specify the time scale over which we shall observe the system.
The CTP action contains this physical time scale $T$. From the
point of view of the RG, this adds one more dimensional parameter
to the theory, much as an external field in the Ising model.
Physically, because time integrations are restricted to the
interval $\left[0,T\right]$, energy conservation does not hold at
each vertex. This is of paramount importance regarding damping.

The RG for the CTP effective  action (obtained by taking the limit
$T\rightarrow\infty$) was studied by Dalvit and Mazzitelli
\cite{dalvit:1996a,dalvit1998}; see also refs. \cite{CHM} and
\cite{Pol06}.  Unlike those works, we focus on the dissipation and
noise features of the effective dynamics, rather than in the
running of the effective potential.

In formulating a nonequilibrium  RG, we must deal with the fact
that the CTP action may have an arbitrary functional dependence on
the fields and be nonlocal both in time and space. In principle,
one can define an exact RG transformation \cite{dalvit:1996a},
where all three functional dependencies are left open. However,
the resulting formalism is too complex to be of practical use.
Fortunately, the special properties of the application to
thermalization allow for a substantial simplifications, such as
working in three spatial dimensions.

The full RG equations for  this theory are given in \cite{us}.
Here we shall only highlight those aspects of the calculation most
relevant to the application to turbulent thermalization.

Let us call $\varphi^{\pm}$  the field variable in the first
(resp. second) branch of the CTP. To write down the CTP action, it
is best to introduce average and difference variables \bea  \sqp
&=& \cp^+ - \cp^-, \\ \cp &=& \cp^+ + \cp^-.\ena In terms of these
variables, a generic CTP action may be written as \beq \fl S_{\rm
CTP}=S_{0}+S_{\lambda}+S_{\rm other}, \enq where $S_{0}$ is the
CTP action functional for a free massless field theory \bea \fl
S_0[\phi, \varphi] =\displaystyle\frac{1}{2}
\displaystyle\int_{0}^{T} \!\!dt \displaystyle\int\! d^d\!k
\;\bigg[ & \dot\phi({{\mathbf k}}, t) \; \dot\varphi(-{{\mathbf
k}}, t) - k^2 \; \phi({{\mathbf k}}, t) \varphi(-{{\mathbf k}},
t)\bigg], \ena $S_{\lambda}$ accounts for a
$\lambda\varphi^{4}$-type self interaction \bea \fl
S_{\lambda}[\sqp, \cp] = &-\D\frac\lambda{48}\D\int_{0}^{T} dt \,
\D\int \D\frac{d^d\!k_1 \dots d^d\!k_4}{(2\pi)^{d}}  \;
\delta^d\lt(\sum_{l=1}^4 \bfk_l\rt) \; \nn \\  \fl \label{the
action 1} & \times \Big[\sqp(\bfk_1, t) \cp(\bfk_2,t)
\cp(\bfk_3,t) \cp(\bfk_4,t)+\sqp(\bfk_1, t) \sqp(\bfk_2,t)
\sqp(\bfk_3,t) \cp(\bfk_4,t)\Big],& \ena and $S_{\rm other}$
includes all other possible terms. Momentum integrals are bounded
by $k = \Lambda$, and $d = 3$. We shall assume that the initial
condition for the RG flow is $S_{\rm other}=0$ at the hard scale
$\Lambda$, so that if it appears at soft scales, it is as a
consequence of the RG running itself. Note that this is true, in
particular, for the noise and dissipation terms.

To define the nonequilibrium RG we also need to specify the state
of the field at the initial time $t=0$. For simplicity, we shall
assume this is the vacuum state for the free action $S_0$. Observe
that this is a nonequilibrium state for the interacting theory.

The value $\lambda_0$ of the coupling constant $\lambda$ at the
hard scale $\Lambda$ may be used as the small parameter in a
perturbative expansion of the RG equation. To order
$\lambda_{0}^{2}$, the RG equation for the quartic coupling
decouples, and can be solved by itself. The result is that at soft
scales $k$, $\lambda$ is both scale and $T$ dependent. There is no
RG running if $T=0$, while the usual textbook result is obtained
as $T\rightarrow\infty$ \cite{Peskin}. For all values of $T$,
$\lambda$ is driven to zero as $k\rightarrow 0$ \cite{us}. Thus it
is consistent to assume that $\lambda$ is uniformly small in the
relevant scale range.

In particular, in order to compute the RG equations to order
$\lambda_0^{2}$, it will be enough to use in the Feynman graphs
the zeroth order propagators, which are those of the massless free
theory. The only exception is in computing the effective mass, but
this calculation is decoupled from the noise and dissipation terms
to order $\lambda_0^{2}$. Observe that it is at the same time a
huge simplification and a strong limitation concerning the range
of application of our results, as we expect substantial shifts in
the propagators when $T$ approaches the relaxation time of the
theory.

Because  of the nonzero initial value of $\lambda$, other
couplings will appear as a result of the RG running. To order
$\lambda_0^{2}$, it is enough to consider quadratic, quartic and
six-point terms in the action. All these terms feed back into each
other, so they must be taken self-consistently. If we understand
thermalization in the usual sense that propagators become
approximately thermal \cite{JCG04}, however, it is enough to focus
on the quadratic terms, \bea S_{\rm other}\rightarrow &S_2[\sqp,
\cp] = \D\int_{0}^{T} \!\! dt_1 \D\int_{0}^{T} \!\! dt_2 \D\int
d^d\!k \; \Big[v_{21}(k;
t_1, t_2) \; \sqp(\bfk, t_1) \, \cp(-\bfk, t_2) & \nn \\ \nn \\
\label{S2} &+ \, {\rm i} \, v_{22}(k; t_1, t_2) \; \sqp(\bfk, t_1)
\, \sqp(-\bfk, t_2) \Big].& \ena

In principle,  the induced quadratic terms will be oscillatory
functions of $\Lambda t_{1,2}$. However, we are interested mostly
in the dynamics of slowly varying field configurations which are
insensitive to high frequencies. To focus on the slow dynamics, we
may project out the mass, dissipation and noise terms on which the
oscillations are mounted.

To this end, we  introduce two projectors. Given a function of two
times $v(k; t_1, t_2)$, we define \bea \label{bfP} \bfP v(k; t_1,
t_2) &=& \cP v(k) \; \delta(t_1-t_2), \ena and, if $v(k; t_1, t_2)
= 0$ for $t_2 > t_1$, \bea \label{bfQ} \bfQ v(k; t_1, t_2) &=& \cQ
v(k) \; \lt[2 \lt(\D\frac{\partial }{\partial t_2} + \delta(t_2) -
\delta(0)\rt) \delta(t_1-t_2)\rt], \ena where \bea \label{def cP}
\cP v(k) &=& \D\frac1T \int_0^T dt_1 \int_0^T dt_2 \; v(k; t_1,
t_2), \ena and \bea \label{def cQ} \cQ v(k) &=& \D\frac1T \int_0^T
dt_1 \int_0^T dt_2 \; v(k; t_1, t_2) \; (t_2-t_1). \ena It is easy
to verify that $\bfP^2 = \bfP$, $\bfQ^2 = \bfQ$, and that $\bfQ
\bfP = \bfP \bfQ = 0$. This proves that the decomposition \bea
\label{decomposition} v(k; t_1, t_2) = \bfP v(k; t_1, t_2) + \bfQ
v(k; t_1, t_2) + \Delta v(k; t_1, t_2) \ena is unique. When this
decomposition is applied to $v_{21}$ in equation (\ref{S2}), we
get \bea \fl \D\int_0^T dt_1 \int_0^T dt_2 \D\int d^d\!k \;
v_{21}(k; t_1, t_2) \;  \sqp(\bfk, t_1) \cp(-\bfk, t_2)   =  \nn \\
\nn \fl \D\int_0^T dt_1 \D\int d^d\!k \left[v_0(k) \; \sqp(\bfk,
t_1) \cp(-\bfk, t_1) + v_1(k)\; \sqp(\bfk, t_1) \dot \cp(-\bfk, t_1) \right]   \nn \\
\fl +\D\int_0^T dt_1 \int_0^T dt_2 \D\int d^d\!k \; \Delta
v_{21}(k; t_1, t_2)\; \sqp(\bfk, t_1) \cp(-\bfk, t_2) , \label{v21
decomposed} \ena where \bea v_0(k) = \cP v_{21}(k) = v_0(0) + k
\D\frac{\partial v_0(0)}{\partial k} + \D\frac{k^2}{2!}
\D\frac{\partial^2 v_0(0)}{\partial k^2} + \ldots,
\label{definicion v0} \ena and \bea \label{definicion v1} v_1(k) =
\cQ v_{21}(k). \ena The linear term in equation (\ref{definicion
v0}) vanishes from symmetry, and the appearance of the quadratic
term is prevented by performing a field rescaling as part of the
RG transformation (thus the field acquires an anomalous
dimension). Neglecting the last term in equation (\ref{v21
decomposed}), the net effect for the long wavelength modes is to
induce a mass term $m^{2} = -2 v_0(0),$ and a damping constant $
\kappa(k) = -v_1(k)$. The noise kernel is obtained in a similar
way from the imaginary part of the CTP action, $v_{22}$ in
equation (\ref{S2}).

After these considerations,   the relevant CTP action for long
wavelength, slowly varying configurations reduces to
\begin{eqnarray} \nonumber  \fl S_{\rm CTP}[\phi, \varphi] =
\displaystyle\int_{0}^{T} \!\!dt \displaystyle\int\! d^d\!k
\;\bigg[ &\frac{1}{2} \dot\phi({{\mathbf k}}, t) \;
\dot\varphi(-{{\mathbf k}}, t) - \displaystyle\frac{1}{2}
\phi({{\mathbf k}}, t)
\left(k^2 + m^2\right) \varphi(-{{\mathbf k}}, t)  \\
\label{the action 2}  &- \kappa(k) \; \phi({{\mathbf k}}, t)
\dot\varphi(-{{\mathbf k}}, t) + \displaystyle\frac{{\rm i}}2
\nu(k) \; \phi({{\mathbf k}}, t) \phi(-{{\mathbf k}},
t)\bigg].\end{eqnarray}

It is shown in Feynman  and Hibbs \cite{if2} that this CTP action
describes a field subject to ohmic dissipation with damping
constant $\kappa$ and a stochastic source $j\left(t\right)$ with
white noise self-correlation $\left\langle
j\left(t\right)j\left(t'\right)\right\rangle =\nu\,
\delta\left(t-t'\right)$. The relationship of the propagators of
the original theory to those obtained from this CTP action is
discussed in \cite{Zanella_talk}. For present purposes, it is
enough to observe that this system thermalizes to an effective
temperature given by the fluctuation-dissipation theorem
\cite{if1, if2,fdt1, fdt2}\begin{equation}
T_{\rm{eff}}=\frac{\nu}{4\kappa},
\end{equation} with a thermalization time \begin{equation}
\tau =\frac{1}{\kappa}.\end{equation} For $k \ll \Lambda$ we
obtain the approximate expressions \cite{us}\begin{eqnarray}
\kappa(k, T) \sim \displaystyle\frac{9}{2T}
\left(\displaystyle\frac{\lambda_0}{96\pi^2}\right)^{\!2}
  \big[7 - 2T^2-8\,\cos T +
\cos(2T)\big] \,\ln\left(\Lambda/k\right), \end{eqnarray} and
\begin{eqnarray} \nonumber \fl \nu\left(k, T\right) \sim
\frac{9}{8 T}
\left(\displaystyle\frac{\lambda_0}{96\pi^2}\right)^{\!2}
\bigg\{\!
135+ 4 \Big[\gamma_{E}- 34\, \ln 2 - 7\, \ln 3+ 7 \,\Ci(3T)\Big]\! - 3 \, \cos(4T) \\
\nonumber \fl + \; 4\,\ln T   - 16\Big[8\pi + 3T - 16\,
\Si(2T)\Big] \sin T + 4\Big[7 \pi + 6 T - 14\, \Si(4T)\Big] \sin (2T) \\
\nonumber \fl -\, 12\Big[19 - 4T^2\Big]\, \Ci(T) +
8\Big[7-6T^2\Big] \Ci(2T) + 4\Big[1-4T^2\Big]\Ci(4T) - 8\,
\cos(3T)\\  \fl - \; 24\big[\,7-8 \, \Ci(2T)\big] \cos T +
4\,\Big[11+2\, \Ci(T) - 2\, \Ci(2T) - 14 \, \Ci(4T)\Big] \cos(2T)
\nonumber  \\  \fl +\;4T\big[ 18\pi  - 54 \,\Si(T)+ 28\,\Si(2T) -
6 \,\Si(3T) - 2\, \Si(4T) +\sin (4T)\big] \bigg\},\end{eqnarray}
where $\Si$ and $\Ci$ are the $\sin$ and the $\cos$ integral
functions. The thermalization time and the effective temperature
go as $\ln(\Lambda/k)^{-1}$ when $k\rightarrow 0$. Observe that
the asymptotic formula for $\kappa$ is not positive definite. This
suggests that the fundamental damping mechanism is Landau damping
of long wavelength waves through interaction with hard quanta
\cite{llpk}. In any case, we expect our approximations to break
down before we reach the point $\kappa = 0$.

In figure 1 we show $\kappa$, $\nu$, and $T_{\rm eff}$ as
functions of the scale $k$ for fixed $T$. In figure 2 we show
$\kappa$, $\nu$, and $T_{\rm eff}$ as  functions of the
observation time $T$ for a fixed  $k$. We have chosen units in
such a way that $\Lambda = 1$. The expressions for $\kappa$ and
$\nu$ are given in \cite{us}.

\begin{figure}
\begin{center}
\includegraphics[angle=0, width = 9cm]{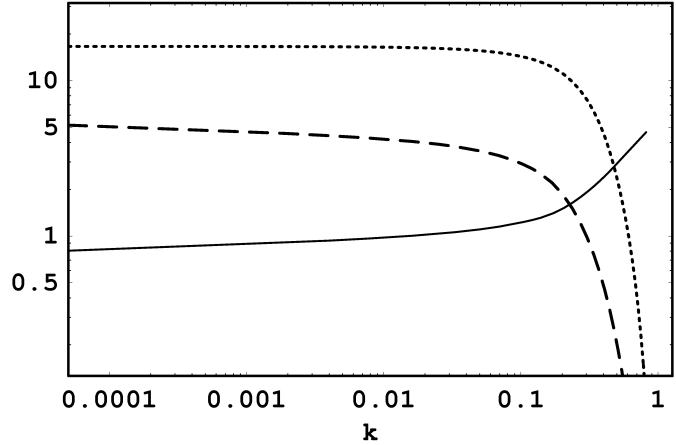}
\caption{The dissipation constant $\kappa$ (dashed), the noise
$\nu$ (dotted), both measured in units of
$({\lambda}/{96\pi^2})^2$, and the effective temperature $T_{\rm
eff}$ as functions of the scale $k$ for a fixed value of $T =
0.5$.}
\end{center}
\end{figure}

\begin{figure}
\begin{center}
\includegraphics[angle=0, width = 9cm]{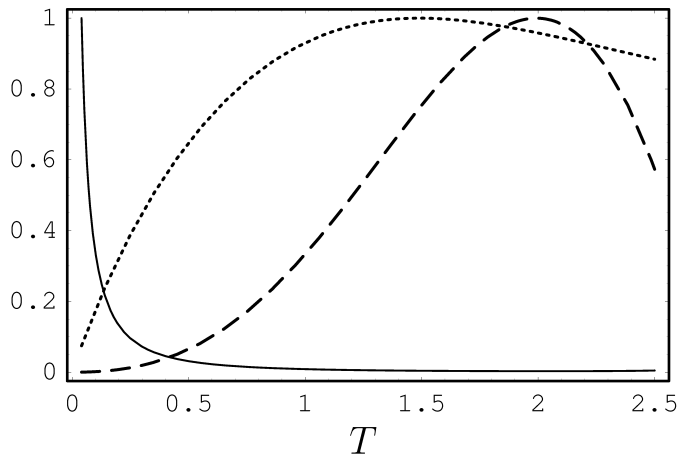}
\caption{The dissipation constant $\kappa$ (dashed), the noise
$\nu$ (dotted), both measured in units of
$({\lambda}/{96\pi^2})^2$, and the effective temperature $T_{\rm
eff}$ (solid), as functions of $T$ for a fixed value of $k =
1/60$. Each quantity is normalized with respect to its maximum
value in the displayed interval($\approx$ 215 for $\kappa$, 26 for
$\nu$ and 10 for $T_{\rm eff}$).}
\end{center}
\end{figure}

The most important result of this paper is that RG running alone
describes the onset of noise and dissipation in the long
wavelength modes, even if these are not assumed to be present in
the underlying microscopic theory. These two elements provide the
sufficient conditions for thermalization as described by Schwinger
\cite{ctp1, ctp2, ctp3, ctp4}. Therefore the model succeeds  in
describing the onset of the thermalization process, and yields
simple estimates of both the final temperature and the
thermalization times at different scales. We must caution however
that in this form these estimates are not fully reliable, as they
involve pushing the theory to the range $\Lambda T\geq 1$. This lies beyond the range of validity of our
approximations concerning $T$. To extend the nonequilibrium RG to
a larger $T$ range a fully self-consistent approach is necessary
\cite{sc1, sc2, sc3}.

Because of this limitation, we do not claim to have solved the
problem, but to have shown a framework for a solution. We need a
self-consistent approach to the hard loops to be able to extend
further the $T$ range. Still, the power of the nonequilibrium RG
allows us to extend to the fully quantum regime the insights
gained from wave turbulence in the classical stage of evolution.

\ack We acknowledge Joan Sola and  Enric Verdaguer for their kind
hospitality at Barcelona, and Enric Verdaguer, Gustavo Lozano,
Daniel Litim, Diego Mazzitelli, Fernando Lombardo and Andreas Ipp
for discussions. This work was supported by Universidad de Buenos
Aires, CONICET and ANPCYT.


\section*{References}

\begin{thebibliography}{99}
\bibitem{Ber04} Berges J 2004 Introduction to Nonequilibrium Quantum Field Theory
{\it Preprint} hep-ph/0409233

\bibitem{BTW06}  Bassett B A, Tsujikawa S and Wands D 2006
{\it Rev. Mod. Phys.} {\bf 78} 537

\bibitem{FeKo06} Felder G  and Kofman L 2006
Nonlinear Inflaton Fragmentation after Preheating {\it Preprint}
hep-ph/0606256.

\bibitem{ZLF92}  Zakharov V E, L'vov V S and Falkovich G 1992  {\it Kolmogorov
spectra of turbulence I: wave turbulence} (Berlin:
Springer-Verlag)

\bibitem{LOMA96}  Lombardo  F and Mazzitelli F D 1996 {\it Phys. Rev.} {\bf D53} 2001

\bibitem{CHM}  Calzetta E, Hu  B L and  Mazzitelli F D 2001 {\it Phys. Rep.} {\bf 352}
459

\bibitem{Pol06} Polonyi J 2006 {\it Phys. Rev.} D {\bf 74} 065014

\bibitem{if1}  Feynman R and  Vernon F 1963 {\it Ann. Phys. (NY)} {\bf{24}} 118

\bibitem{if2} Feynman R and Hibbs A 1965 {\it Quantum Mechanics and Path Integrals}
(New York: McGraw - Hill)

\bibitem{WiKo74}  Wilson K and Kogut J 1974 {\it Phys. Rep.} {\bf 12} 75

\bibitem{BOD98} Bodeker D 1998 {\it Phys. Lett.} B {\bf 426} 351

\bibitem{BOD99} Bodeker D 1999  {\it Nuc.
Phys.} B {\bf 559} 502

\bibitem{LIMA02}  Litim D and  Manuel C 2002 {\it Phys. Rep.} {\bf 364} 451

\bibitem{LIT98} Litim D 1998 Wilsonian flow equations and thermal field theory {\it Preprint}  hep-ph/9811272

\bibitem{ctp1}  Schwinger J 1961 {\it J. Math. Phys.} {\bf 2} 407

\bibitem{ctp2} Keldysh L V 1964  {\it Zh. Eksp. Teor. Fiz.} {\bf{47}} 1515
[Engl. trans: Keldysh L V 1965 {\it Sov. Phys.} JEPT {\bf{20}}
1018 ]

\bibitem{ctp3} Calzetta E and Hu B L 1987 {\it Phys. Rev.} D {\bf{35}} 495

\bibitem{ctp4}  Calzetta E and Hu B L 1988 {\it Phys. Rev.} D {\bf{37}} 2878

\bibitem{dalvit:1996a} Dalvit D A R and Mazzitelli F D 1996 {\it Phys. Rev.} D {\bf 54}
6338

\bibitem{dalvit1998} Dalvit D A R 1998 Ph D Thesis (Universidad de Buenos
Aires)

\bibitem{us} Zanella J and Calzetta E 2006 Renormalization group study of damping in nonequilibrium field
theory {\it Preprint} hep-th/0611222

\bibitem{Peskin} Peskin M E and Schroeder D V 1995 {\it An Introduction to Quantum
Field Theory} (Perseus Books, Cambridge, Massachusetts)

\bibitem{JCG04}  Juchem S, Cassing W and Greiner C 2004 {\it Phys. Rev.} D {\bf 69}
025006


\bibitem{Zanella_talk} Zanella J and Calzetta E 2006 Inflation and nonequilibrium renormalization group {\it Preprint} hep-ph/0611335

\bibitem{fdt1} Callen H and Welton T 1951 {\it Phys. Rev.} {\bf 83}
34

\bibitem{fdt2} Landau L, Lifshitz E M and Pitaevsky L 1980 {\it Statistical Physics}
vol 1 (London: Pergamon)

\bibitem{llpk}  Lifshitz E M and Pitaievskii L P 1981 {\it Physical
Kinetics} (Oxford: Pergamon Press)

\bibitem{sc1} Salmhofer M 2006 Dynamical Adjustment of Propagators in Renormalization Group Flows
{\it Preprint} cond-mat/0607289

\bibitem{sc2} Litim D F and  Pawlowsky J M 2006 Non-perturbative thermal flows and resummations
{\it Preprint} hep-th/0609122

\bibitem{sc3} Blaizot J-P, Ipp A, Mendez-Galain R and
Wschebor N 2006 Perturbation theory and non-perturbative
renormalization flow in scalar field theory at finite temperature
{\it Preprint} hep-ph/0610004

\end{thebibliography}
\end{document}